\documentclass[12pt]{article}
\usepackage[a4paper,left=30mm,right=20mm,top=20mm,bottom=20mm]{geometry}
\usepackage{graphics}
\usepackage{amsmath}
\usepackage{amssymb}
\usepackage{epsfig}
\usepackage{cite}
\usepackage{xcolor}
\usepackage[english]{babel}
\def\MGvATNLO{{\sc MadGraph5\_aMC@NLO}}

\title{Probing the proton structure with associated vector boson and heavy flavor jet production at the LHC}

\author{A.V.~Lipatov$^{1,2}$, G.I.~Lykasov$^2$, M.A.~Malyshev$^{1}$, S.M.~Turchikhin$^2$}

\begin{document}

\maketitle

\begin{center}

{\it $^{1}$Skobeltsyn Institute of Nuclear Physics, Lomonosov Moscow State University, 119991, Moscow, Russia}\\
{\it $^{2}$Joint Institute for Nuclear Research, 141980, Dubna, Moscow region, Russia}\\

\end{center}

\vspace{0.5cm}

\begin{center}

{\bf Abstract }

\end{center}

\indent

We consider the production of $Z$ bosons associated with heavy (charm and beauty) 
jets at the LHC energies using two scenarios based on the transverse momentum dependent (TMD)
parton densities in a proton. The first of them employs the 
Catani-Ciafaloni-Fiorani-Marchesini gluon evolution and 
is implemented in the Monte-Carlo event generator \textsc{pegasus}.
Here, the heavy quarks are always produced in the hard partonic scattering.
The second scheme is based on the parton branching approach, currently
implemented into the Monte-Carlo event generator \textsc{cascade}.
In this scenario, the $Z$ + jets sample is generated and then events 
containing the heavy flavor jet in a final state are selected.
We compare the predictions obtained within these two TMD-based approaches to each other,
investigate their sensitivity to the TMD gluon densities in a proton
and estimate the effects coming from parton showers and double parton scattering mechanism.
Additionally, we compare our predictions
with the results of traditional (collinear) pQCD calculations
performed at NLO accuracy.
It is shown that the TMD-based results agree with the LHC 
experimental data collected at $\sqrt s = 8$ and $13$~TeV.
We discuss the sensitivity of observables to the quark 
distributions in a proton
and present predictions to search for the intrinsic charm signal in 
forthcoming analyses of the LHC experimental data.

\vspace{1.0cm}

\noindent{\it Keywords:} QCD evolution, small-$x$, TMD gluon densities in a proton, electroweak bosons, heavy quarks

\newpage

\section{Introduction} \indent

Recently, the ATLAS and CMS Collaborations have presented measurements
\cite{1,2,3,4,5,6,7} of the
total and differential cross sections of $Z$ boson and associated 
heavy (charm and beauty) quark jet(s) production in $pp$ collisions at the LHC.
These processes are the so-called "rare" processes which
provide a test of the perturbative Quantum 
Chromodynamics (pQCD) predictions and which
could have never been 
systematically studied at previous accelerators.
A good description of the $Z$ boson and heavy flavor jet production
is important since it is one of major background for a variety of 
physics processes, for example, 
associated $Z$ and Higgs boson production.
Moreover, it
can be used to search for new physics 
signatures
and to investigate the quark and gluon content of a proton.
In particular, $Z + c$ events can be used
to study the possibility of observing an intrinsic charm (IC) 
component~\cite{Beauchemin:2014,Lipatov:2016}. The existence of such contribution was originally proposed 
in the BHPS model\cite{8} and developed further in subsequent
papers\cite{9,10} (see also recent review\cite{11}).

The reported measurements\cite{1,2,3,4,5,6,7} are found to be in good 
agreement with the next-to-leading-order (NLO) pQCD 
predictions\cite{12,13,14} based on the four-flavor (4FS)
and five-flavor (5FS) schemes\footnote{The discussion on the advantages 
and disadvantages of the different flavor number schemes can be found, for
example, in review\cite{15}.}.
These predictions were obtained
using \textsc{mcfm}\cite{12}, \textsc{mg5}\_a\textsc{mc}\cite{13}
and \textsc{sherpa}\cite{14} packages,
where the hard scattering processes were simulated and
combined with parton showering and hadronization procedures.
The different jet multiplicities are combined at the 
amplitude level and merged, for example, with the 
FxFx\cite{16} or MiNLO methods\cite{17}. 
Despite the fact that the developed approaches for matching and merging matrix element 
evaluations and parton showers are rather successful, several 
points essential at high energy collisions are not fully treated. 
First, the hard scattering 
amplitudes are calculated within the collinear dynamics and inclusion of the
initial state parton showers results in a net transverse momentum of the hard 
process. Second, 
the special treatment of high energy effects 
is not included.

An alternative description of the LHC data\cite{1,2,3,4,5,6,7} can
be achieved in the
framework of approaches\cite{18,19,20} which involve 
the high-energy QCD factorization\cite{21},
or $k_T$-factorization\cite{22} prescription.
The latter is mainly based on the Balitsky-Fadin-Kuraev-Lipatov (BFKL)\cite{23} or
Catani-Ciafaloni-Fiorani-Marchesini (CCFM)\cite{24} evolution
equations, which resum large terms proportional 
to $\alpha_s^n \ln^n s/\Lambda_{\rm QCD}^2 \sim \alpha_s^n \ln^n 1/x$,
important at high energies\footnote{The CCFM equation additionally 
takes into account terms proportional to $\alpha_s^n \ln^n 1/(1 - x)$ and 
therefore can be applied for both small and large $x$\cite{24}.}
$s$ (or, equivalently, at small $x \sim \mu/\sqrt s$, where $\mu$ is the 
typical hard scale of the process under consideration).
The $k_T$-factorization approach has certain technical advantages in the ease of
including higher-order pQCD radiative corrections (namely, dominant parts of NLO + NNLO + ... terms
corresponding to real initial-state gluon emissions) in the form of
transverse momentum dependent (TMD) parton densities in a proton\footnote{The detailed 
description of this approach can be found, for example, in review\cite{25}.}.
Early calculations\cite{18,19} performed in a "combined" scheme employing both 
the $k_T$-factorization and conventional (collinear) QCD factorization, with 
each of them used in the kinematic conditions of its best reliability,
show reasonably good agreement with the first LHC data 
for $Z + b$ production collected at $\sqrt s = 7$~TeV.
A more rigorous consideration\cite{20}, based on the
Parton Branching (PB) approach\cite{26,27},
leads to similar results.
The PB approach provides an iterative solution of the 
Dokshitzer-Gribov-Lipatov-Altarelli-Parisi
(DGLAP) evolution equations for conventional and TMD quark and gluon densities 
in a proton. The main advantage of the PB scenario is that the TMD parton
densities (and all corresponding non-perturbative parameters)
can be fitted to experimental data, so that the relevant 
theoretical predictions, where the parton shower 
effects are already taken into account, 
can be obtained with no further free parameters --- that 
is in contrast to the usual parton shower event generators\footnote{The 
correspondence between the CCFM and PB based scenarios has been established\cite{20}.}.

In the present paper we improve the early calculations\cite{18,19}
of associated $Z$ boson and heavy flavor jet production
by including into the consideration the effects of 
parton showers in the initial and final states
and extend them to the latest LHC data on $Z + c$-jet production 
collected at $\sqrt s = 8$ and $13$~TeV\cite{2,4,6}.
The predictions, based mainly on the CCFM 
gluon dynamics in a proton, will be compared with the 
results obtained in the PB scenario, 
implemented in the Monte-Carlo event generator \textsc{cascade}\cite{27}.
Such comparison between the calculations performed within these two 
approaches could be a general consistency check for the $k_T$-factorization 
phenomenology.
At this point, our study is complementary to recent investigations\cite{19, JungZb2021}.
Special interest is related to the 
comparison of the TMD-based predictions and traditional (collinear) pQCD ones 
calculated by taking into account higher-order terms.
We consider predictions from the standard \textsc{MadGraph5}\_a\textsc{mc@nlo} tool\cite{madgraph}.
Another goal is connected
with studying the heavy quark density functions in a proton, which 
is particularly interesting for the analysis of hard
processes at LHC energies.
We investigate the influence
of IC contributions on various kinematical distributions in $Z + c$-jet 
production (and, of course, on the recently measured $\sigma(Z + c)/\sigma(Z + b)$ 
relative production rates). We describe new observables 
which are sensitive to the IC content of a proton. 
In this sense we continue the line of our previous studies\cite{28,29}.
Finally, we investigate the role of an additional mechanism of $Z + c$ production,  
double parton scattering (DPS), which is widely discussed in the 
literature at present (see, for example,\cite{30,31,32,33,34,35,36,37,SnigRev} and references therein).

The outline of our paper is following. In Section~2 we briefly describe
our theoretical input. The numerical results and discussion 
are presented in Section~3. Our conclusions are summarized in Section~4.

\section{Theoretical framework} \indent

In the present paper to calculate the total and differential cross sections of associated $Z$ boson and heavy flavor jet 
production at LHC conditions we apply two schemes based on the $k_T$-factorization formalism,
which can be considered as a convenient alternative to higher-order DGLAP-based calculations.
The first scheme was proposed in\cite{18}
and relies mainly on the ${\cal O}(\alpha \alpha_s^2)$ off-shell (depending on the 
transverse momenta of initial particles) gluon-gluon fusion subprocess:
\begin{gather}
  g^* + g^* \to Z + Q + \bar Q,
\label{ggZ}
\end{gather}
\noindent
which gives the leading contribution to the production cross section
in the small $x$ region, where the gluon density dominates over the quark distributions.
An essential point here is using the CCFM evolution equation to
describe the QCD evolution of the transverse momentum dependent (TMD) gluon density in a proton (see\cite{25}).
This equation smoothly interpolates between the small-$x$ BFKL gluon dynamics 
and high-$x$ DGLAP one, thus 
providing us with a suitable tool for the phenomenological study.
In addition to that, we take into account several subleading
subprocesses involving quarks in the initial state --- flavor excitation subprocess
\begin{gather}
  q + Q \to Z + Q + q,
\label{qqtZ}
\end{gather}
\noindent
quark-antiquark annihilation subprocess 
\begin{gather}
  q + \bar q \to Z + Q +\bar Q,
\label{qqsZ}
\end{gather}
\noindent
and quark-gluon scattering
\begin{gather}
  q + g \to Z + q + Q + \bar Q,
\label{qg5}
\end{gather}
\noindent
which could play a role at large transverse 
momenta (or, respectively, at large $x$)
where quarks are less suppressed or can even dominate over the gluon density.
The last subprocess is taken into account since it provides additional heavy quarks, 
despite they are obviously suppressed in strong coupling $\alpha_s$.
Thus, taking into account the subprocesses~(\ref{qqtZ}) --- (\ref{qg5}) 
extends the predictions to the whole kinematic range.
Note that one has at least one heavy quark $Q$ in the final state already at the amplitude 
level.

The gauge-invariant off-shell amplitude for subprocess~(\ref{ggZ}) was 
calculated earlier\cite{ggZbb1,ggZbb2}, where all details are explained.
In contrast with the off-shell gluon-gluon fusion,
the contributions from quark-involved 
subprocesses~(\ref{qqtZ}) --- (\ref{qg5}) are taken into
account using the DGLAP-based factorization scheme, which provides 
better theoretical grounds in the region of large $x$.
The evaluation of the corresponding production amplitudes is straightforward and needs no explanation. 
We only note that the subsequent decay $Z \to l^+l^-$ (including the $Z/\gamma^*$ interference 
effects) is incorporated already at the production step at the amplitude level in order to 
fully reproduce the experimental setup.
To calculate the 
contribution from the off-shell gluon-gluon fusion subprocess~(\ref{ggZ})
we used two latest sets\footnote{A comprehensive collection of TMD gluon
densities can be found in the \textsc{tmdlib} package\cite{tmdlib}, which is a C++ library 
providing a framework and interface to different parametrizations.} of CCFM-evolved 
TMD gluon densities in a proton, namely, JH'2013 set 1 and set 2\cite{Hautmann:2013}.
Their input parameters have been derived from a description of high precision HERA data 
on proton structure functions $F_2(x,Q^2)$ and/or $F_2^c(x,Q^2)$. 
For quark-induced subprocesses~(\ref{qqtZ}) --- (\ref{qg5}) 
we have applied the standard CT14 (NNLO) set\cite{CT14NNLOIC}.

The scheme\cite{18} represents a combination of two techniques
with each of them being used at the kinematic conditions where it is best 
suitable. This scheme is implemented into the Monte-Carlo event generator \textsc{pegasus}\cite{pegasus},
which has been used in the numerical calculations below.
Additionally, we simulate here the effects of parton showers in the 
initial and final states using the \textsc{pythia8}\cite{pythia}, thus
improving the previous consideration\footnote{The TMD parton shower tool
implemented into the Monte Carlo generator \textsc{cascade}\cite{27} is applied for 
off-shell gluon-gluon fusion subprocess~(\ref{ggZ}).}\cite{18,19}.
The resulting partons are then processed with \textsc{fastjet}\cite{fastjet} 
to reconstruct jets in anti-$k_T$ algorithm with radia $R_{\rm jet}$ corresponding to 
the experimental setup. As the heavy quark jet we take the jet, which passes 
kinematical cuts of the experiment and in which a heavy quark is situated closest 
to the jet axis in $\Delta R=\sqrt{\Delta\eta^2+\Delta\phi^2}$,
where $\Delta \eta$ and $\Delta\phi$ are the corresponding differences 
in pseudo-rapidity and azimuthal angle.
In order to avoid the double counting the during parton shower simulation,
we keep the subprocess~(\ref{qg5}) at parton level calculations only.

We compare our results with a more rigorous scheme
based on the Parton Branching (PB) approach\cite{26,27}, which provides 
a solution of the DGLAP equations
for conventional and TMD quark and gluon distributions in a proton.
The splitting kinematics at each branching vertex is described by the DGLAP equations. 
Instead of the usual DGLAP ordering in virtuality, angular ordering condition for parton emissions 
is applied. 
One of the advantages of this approach is that the PB TMDs can be combined with standard (on-shell) production amplitudes, which can be calculated at higher orders. Here we use matrix elements calculated with next-to-leading (NLO) order with  \MGvATNLO\  \cite{madgraph} using the HERWIG6 subtraction terms, which are suitable for combination with PB-TMDs.
A special procedure is adopted for
the transverse momenta of initial partons: a transverse momentum is
assigned according to the TMD density, and then the parton-parton system is boosted 
to its center-of-mass frame and rotated in such a way that only the longitudinal and 
energy components are nonzero. The energy and longitudinal component of the initial momenta 
are recalculated taking into account the virtual masses\cite{27}. 
This method keeps the parton-parton invariant mass exactly conserved, while 
the rapidity of the partonic system is approximately restored.

Similar to the CCFM scenario, the PB TMD parton densities can be obtained via fitting to precise DIS data. 
Two sets, which differ from each other by a choice of the scale in QCD
coupling, were obtained in Ref.~\cite{38}. In the numerical calculations below 
we have used the PB-NLO-HERAI+II-2018 set 2. 
Technically, we generate a $Z$ + jet(s) sample using \textsc{cascade}
and then select events which contain the heavy flavor jet(s) in a final 
state (see also\cite{JungZb2021}).
This is in contrast to the \textsc{pegasus} calculations, 
where a heavy flavor jet is always presented in the final state, as explained above.

Finally, we turn to the DPS contribution to $Z + c$ production.
We apply the factorization formula\cite{30,31,32,33,34,35,36,37,SnigRev}:
\begin{gather}
  \sigma_{\rm DPS}(Z + c) = {\sigma(Z) \sigma(c) \over \sigma_{\rm eff}}, 
\end{gather}
\noindent 
where $\sigma_{\rm eff}$
is a normalization constant which incorporates all "DPS unknown" into a single 
phenomenological parameter.
A numerical value $\sigma_{\rm eff} \simeq 15$~mb
was obtained, for example, in recent studies\cite{69,70,71,72,73} from fits to Tevatron and LHC data (see also\cite{74}). 
This will be taken as the default value throughout the paper.
The calculation of inclusive $Z$ boson or charm production cross sections 
is straightforward and needs no special explanations.
Here we strictly follow the approach described earlier\cite{75,76,77}.

\section{Numerical results} \indent

\begin{table} \footnotesize 
\label{table1}
\begin{center}
\begin{tabular}{|c|c|c|c|c|}
\hline
 & CMS   & CMS & CMS & $x_F$ calculation \\
 & {\footnotesize $\sqrt{s}=8$~TeV \cite{5}}  & {\footnotesize $\sqrt{s}=13$~TeV \cite{6}}  & {\footnotesize $\sqrt{s}=13$~TeV \cite{7}} &  \\\hline
ordered $p_T^l$, GeV & $>20, 20$ & $>26, 10$ & $>25, 25$ &  $>28, 28$  \\\hline
$|\eta^l|$ & $<2.1$ & $<2.4$  & $<2.4$ & $<2.5$  \\\hline
$m^Z$, GeV & \multicolumn{4}{c|}{71---111}  \\\hline
lepton isolation $\Delta R$ & 0.5 & 0.4 & 0.3 (0.4) & 0.4  \\\hline
$R_{\text {jet}}$ & 0.5 & 0.4 & 0.4 & 0.4 \\\hline
$p_T^{\text {jet}}$, GeV & $>25$ & $>30$ & $>30$ & $>20$ \\\hline
$|\eta^{\text {jet}}|$ & $<2.5$ & $<2.4$ & $<2.4$ & $<2.5$ \\\hline
\end{tabular}
\end{center}
\caption{Basic parameters, used for simulations of associated $Z + c$-jet production. By default experimental 
cuts for electrons are shown. Cuts for muons are placed in brackets, if differ.}
\end{table}

Before presenting results of our calculations let us describe our 
set of parameters. 
So, following\cite{PDG}, we apply charm and beauty quark masses
$m_c = 1.4$~GeV and $m_b = 4.75$~GeV, mass of $Z$ boson $m_Z = 91.1876$~GeV, its
total decay width $\Gamma_Z = 2.4952$~GeV and $\sin^2\theta_W = 0.23122$.
As it was mentioned above, we kept $n_f = 4$ active (massless) quark flavors 
in the \textsc{pegasus} calculations, set 
$\Lambda_{\rm QCD}^{(4)} = 200$~MeV and used two-loop QCD coupling according 
to\cite{Hautmann:2013}.
The default renormalization scale was taken to be $\mu_R^2 = m_Z^2$. 
The default factorization scale for the off-shell gluon-gluon fusion 
subprocess was taken as $\mu_F^2 = \hat s + {\mathbf Q}_T^2$,
where ${\mathbf Q}_T$ is the net transverse momentum of the initial
off-shell gluon pair. This choice
is dictated mainly by the CCFM evolution algorithm (see\cite{Hautmann:2013} for more information). 
For quark-induced 
subprocesses~(\ref{qqtZ}) and (\ref{qqsZ}) we keep it equal to the renormalization scale.

The PB calculation with CASCADE~\cite{Jungprivate} were calclualted with $m_c = 1.47$~GeV, $m_b = 4.5$~GeV, $\alpha_s (m_Z^2) = 0.118$ and $\mu_R = \mu_F = \frac{1}{2} \sum_i \sqrt{m^2_i +p^2 _{t,i}}$,  where the sum runs over all particles and parton in the matrix element.  The hard process calculations are performed at NLO with \MGvATNLO\  ~\cite{madgraph} with  \textsc{herwig6} subtraction terms. The theoretical uncertainties are obtained by varying the scale of the hard process by a factor 2 up and down, provided by \MGvATNLO .


\begin{figure}
\begin{center}
\includegraphics[width=7.8cm]{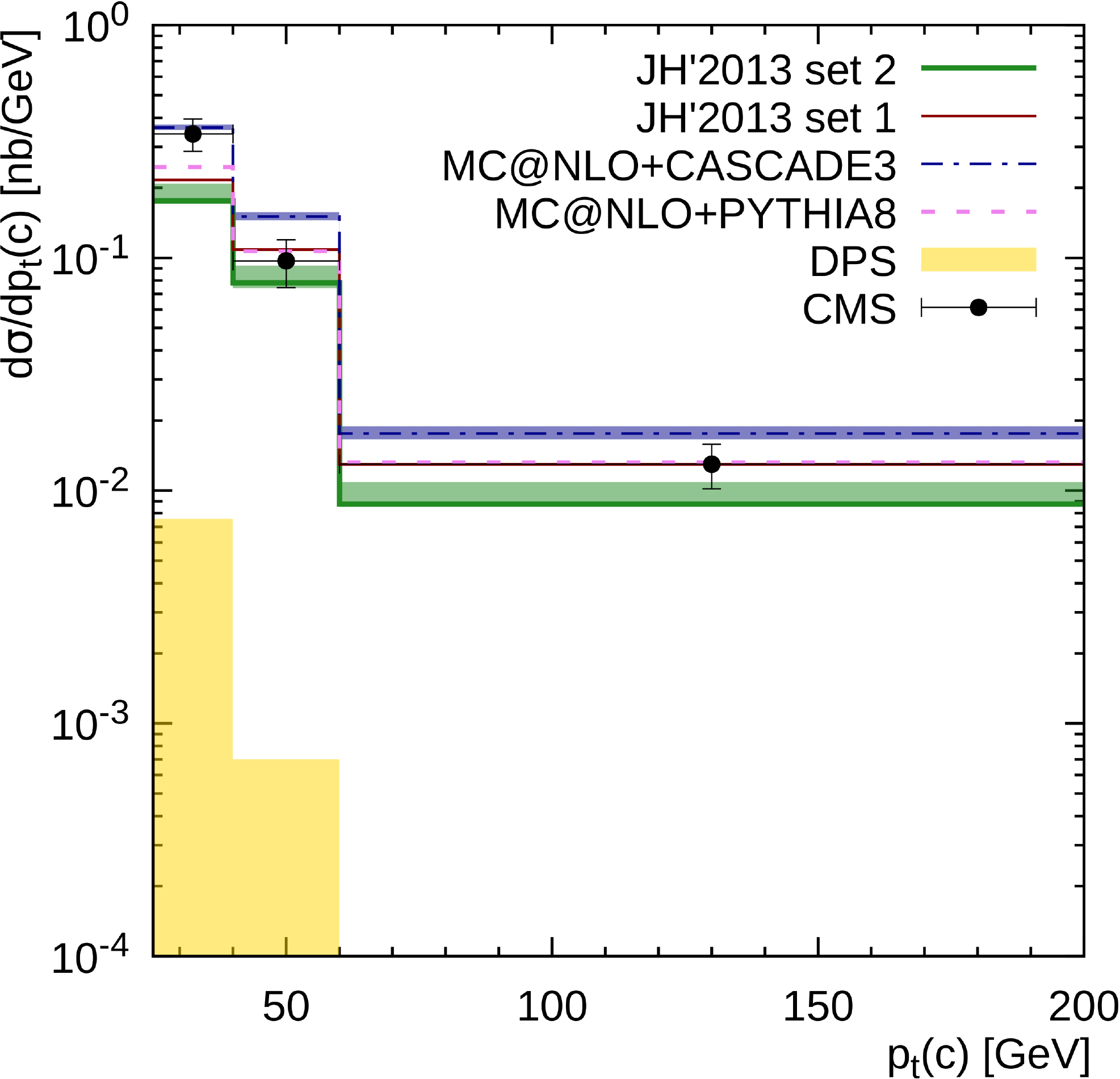}
\includegraphics[width=7.8cm]{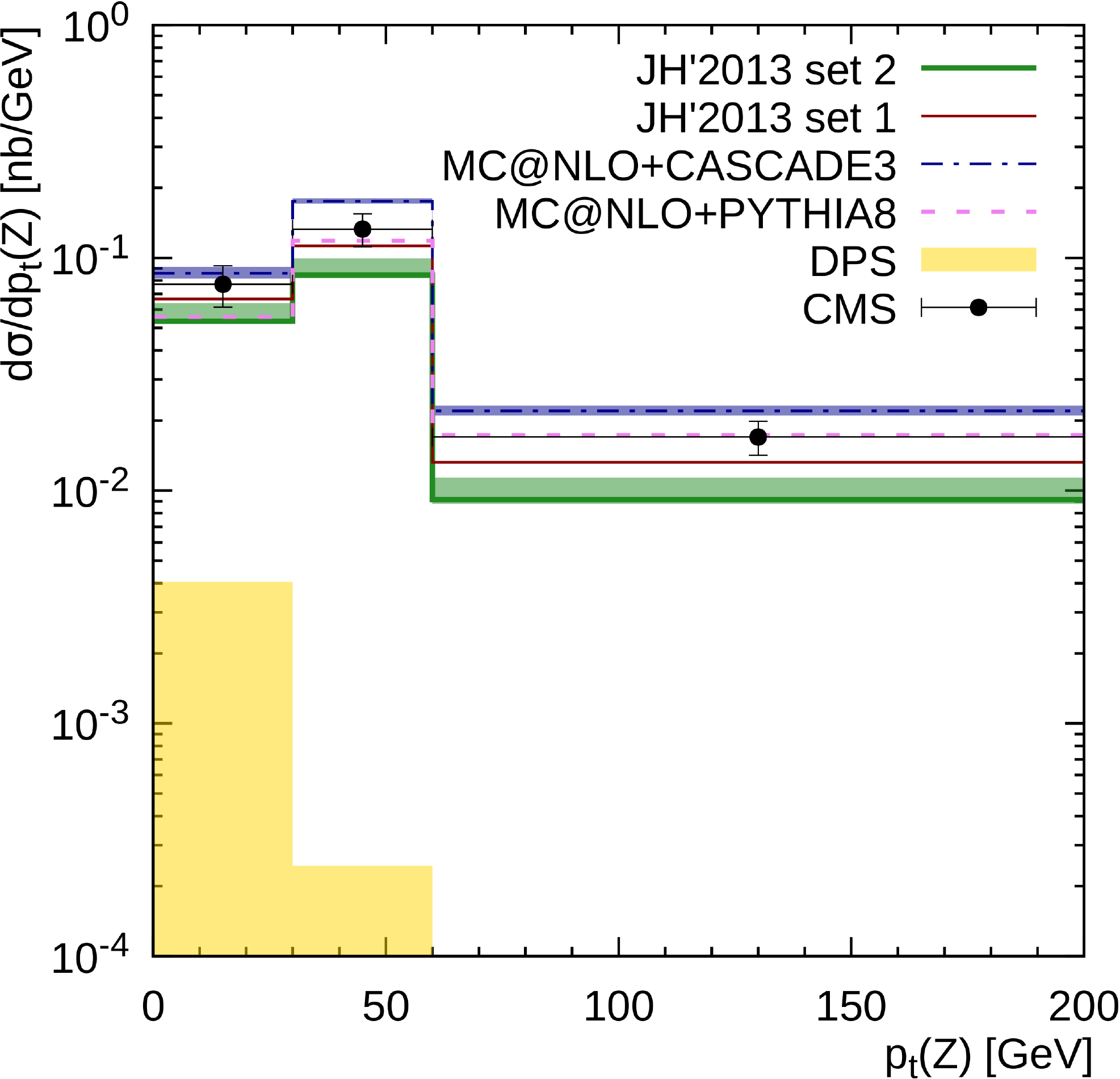}
\includegraphics[width=7.8cm]{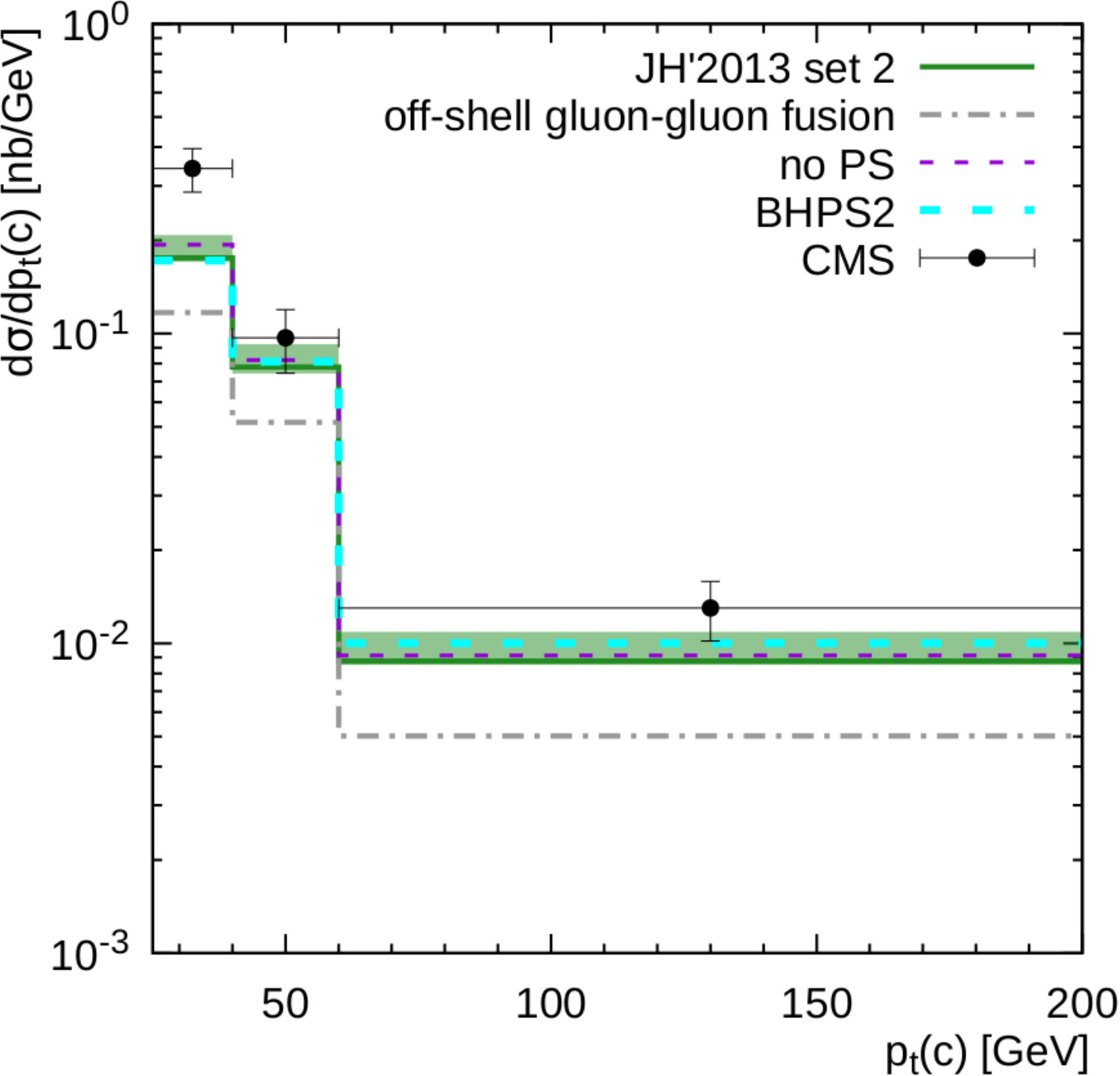}
\includegraphics[width=7.8cm]{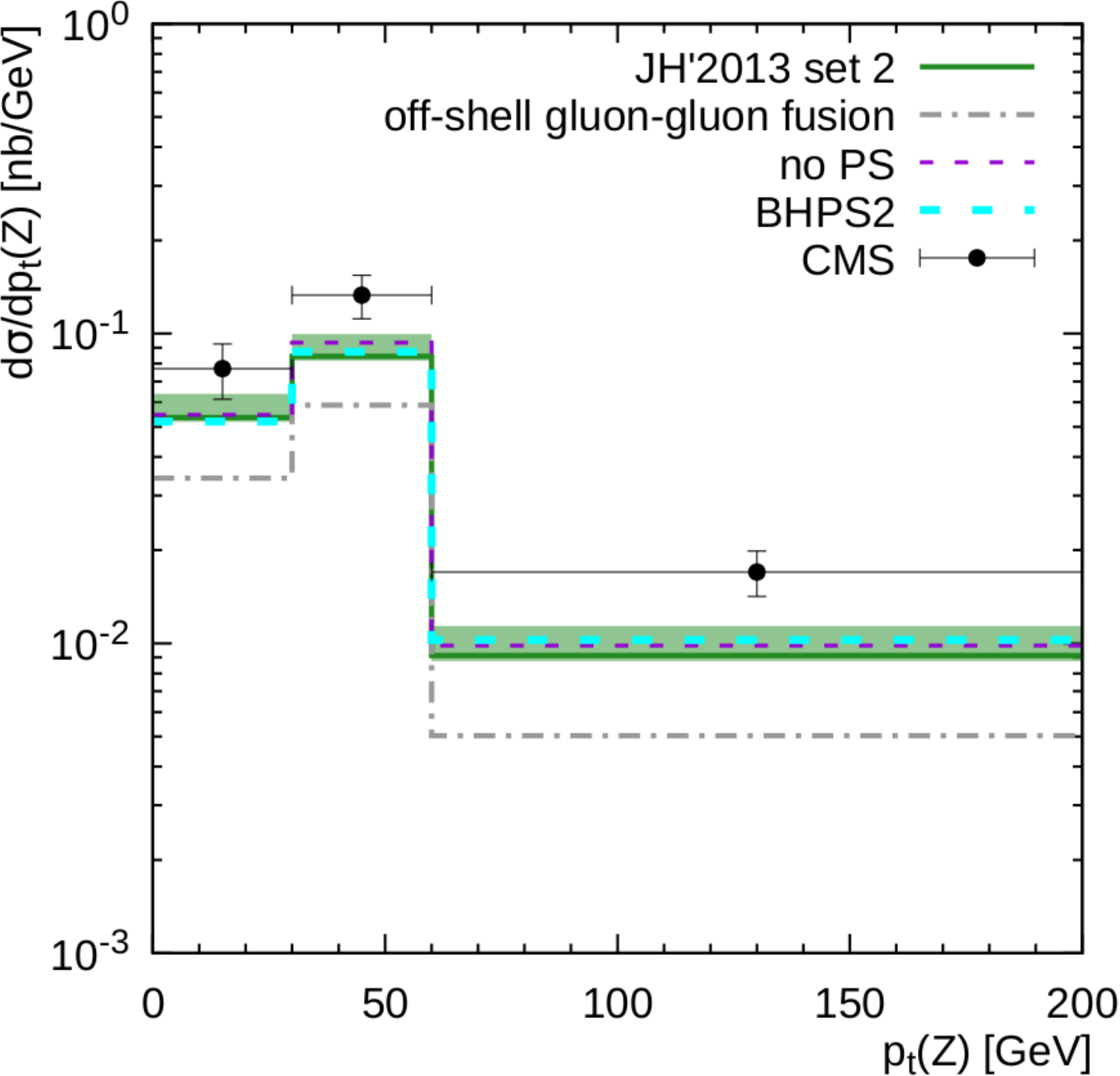}
\caption{The differential cross sections of $Z + c$-jet 
production in $pp$ collisions calculated as functions of 
$c$-jet (left panel) and $Z$ boson (right panel) transverse momenta at $\sqrt{s} = 8$~TeV. 
Shaded bands represent the theoretical uncertainties of our 
calculations, estimated as explained in the text.
The contributions from off-shell gluon-gluon fusion subprocess
are shown separately. The roles of parton showers,
DPS production mechanism and IC terms are illustrated also.
The experimental data are from CMS\cite{5}.}
\label{fig1}
\end{center}
\end{figure}

\begin{figure}
\begin{center}
\includegraphics[width=7.8cm]{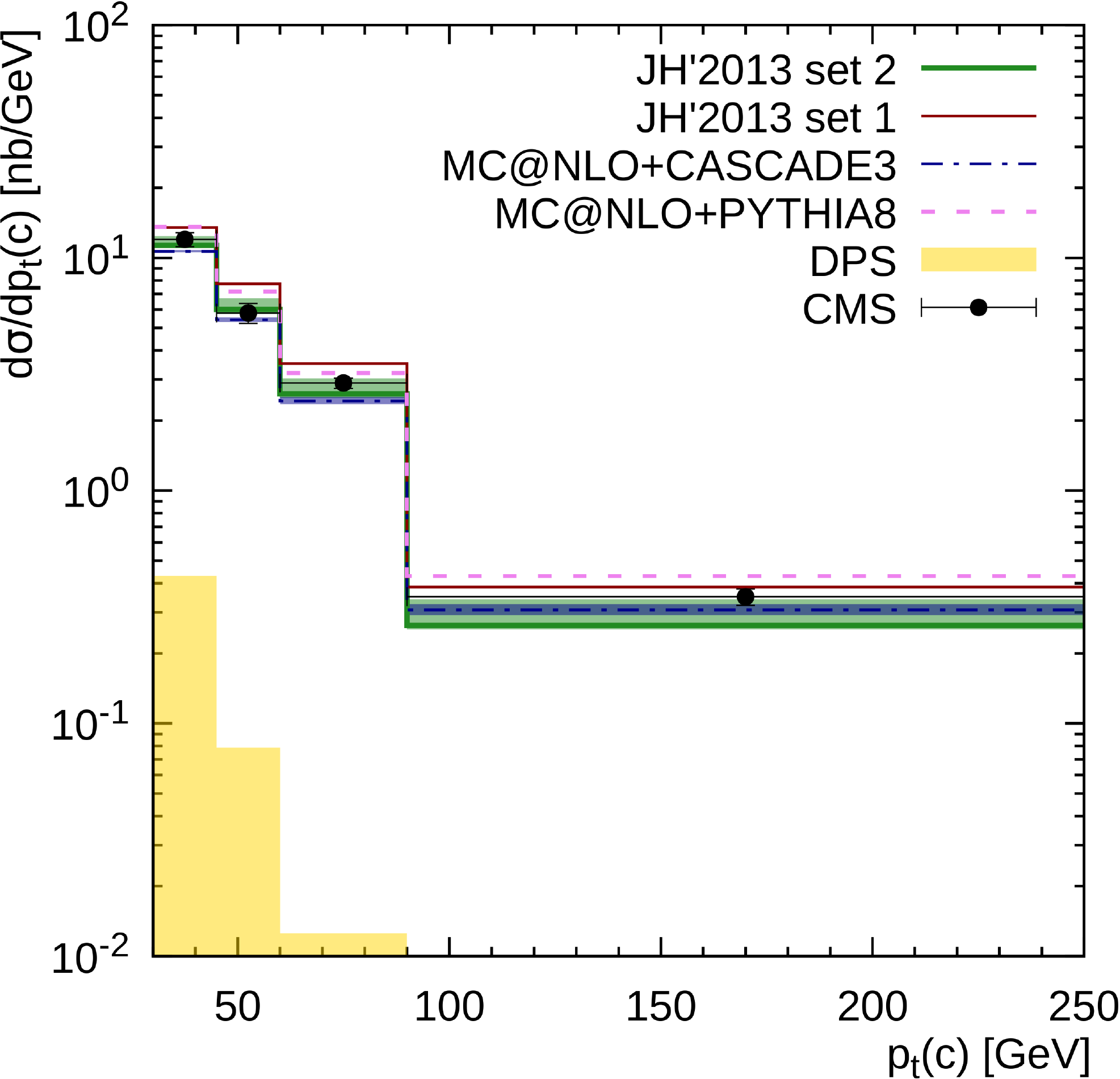}
\includegraphics[width=7.8cm]{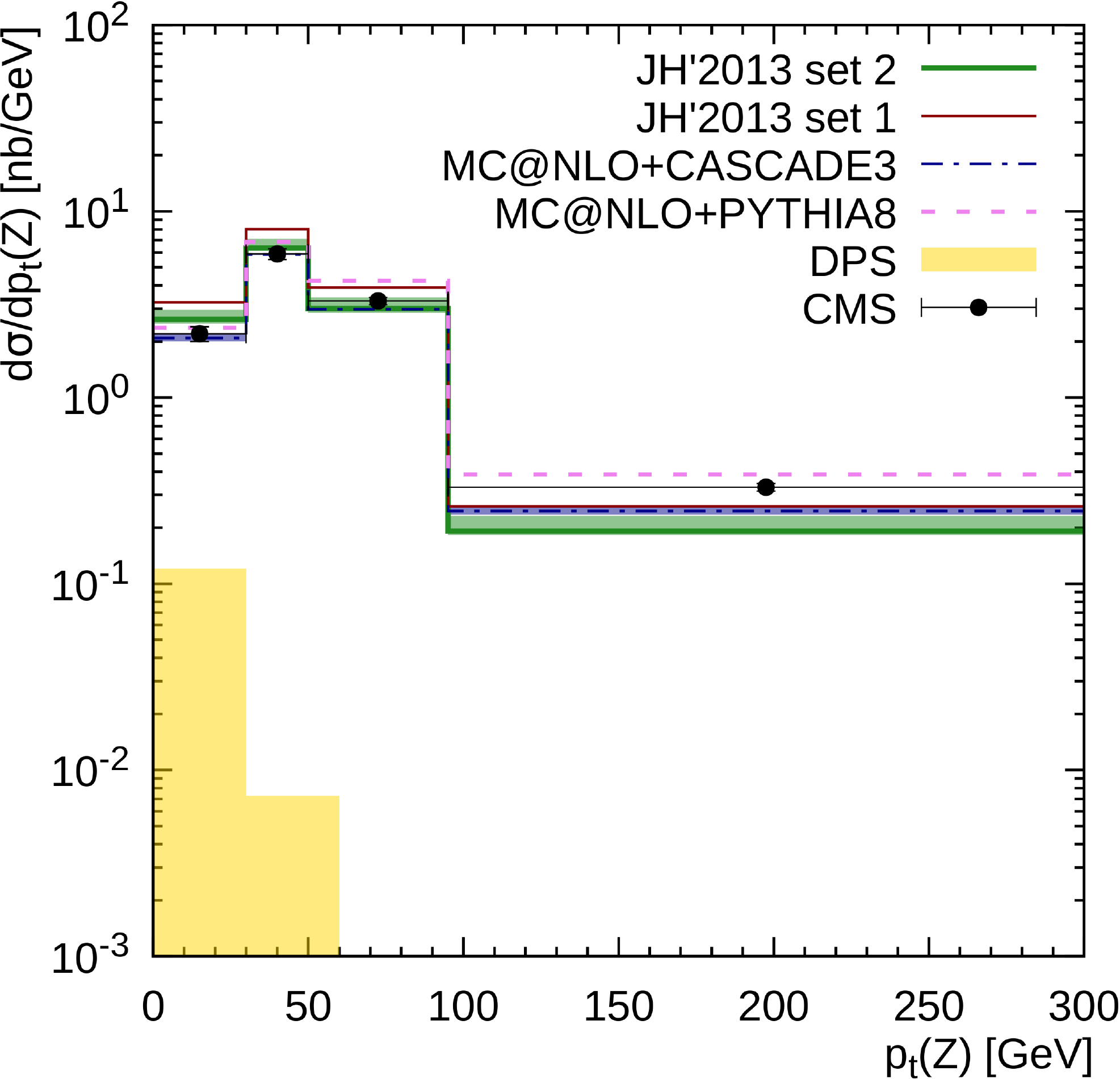}
\includegraphics[width=7.8cm]{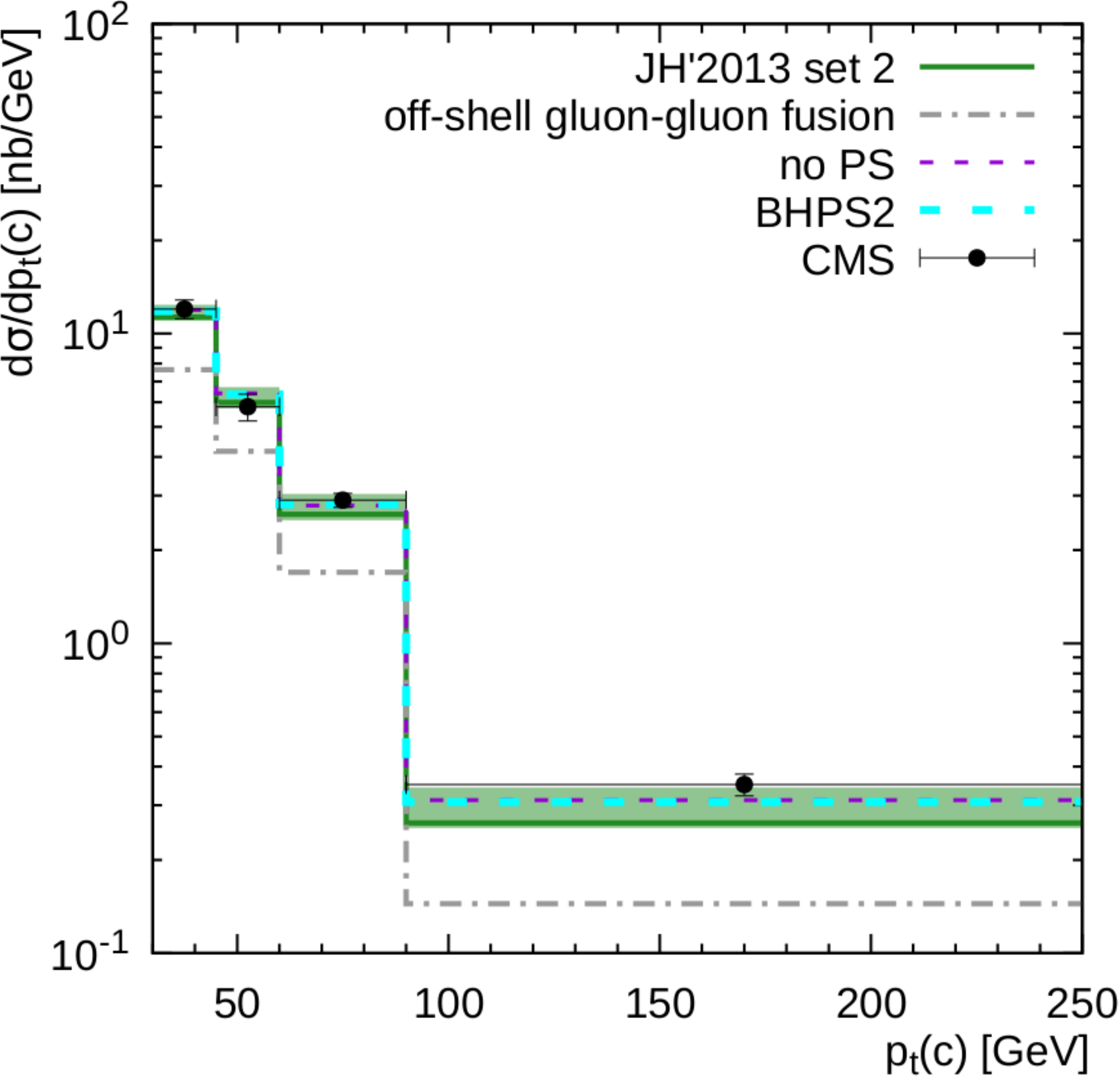}
\includegraphics[width=7.8cm]{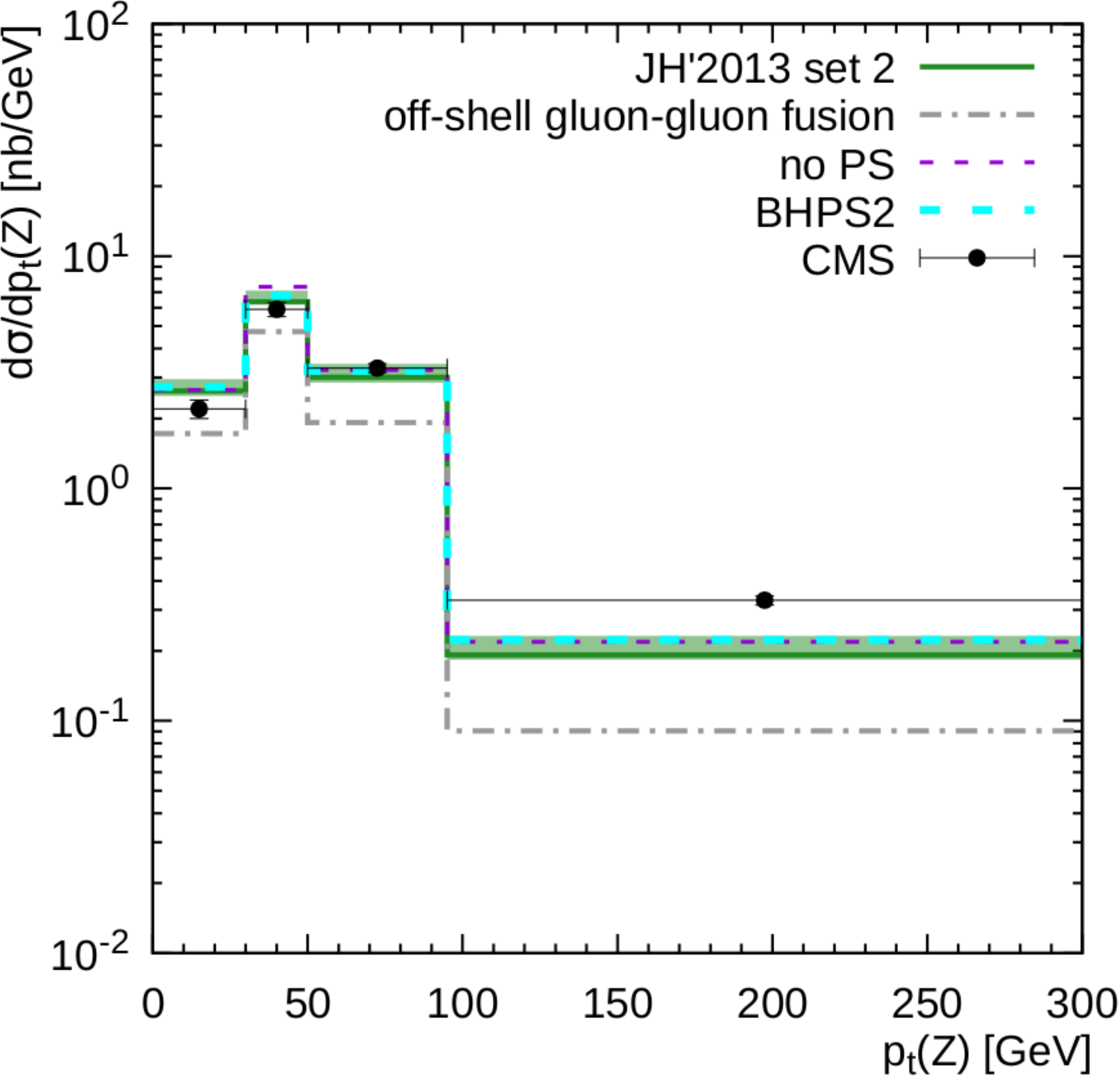}
\caption{The differential cross sections of $Z + c$-jet 
production in $pp$ collisions calculated as functions of 
$c$-jet (left panel) and $Z$ boson (right panel) transverse momenta at $\sqrt{s} = 13$~TeV. 
The notations are the same as in Fig.~1.
The experimental data are from CMS\cite{6}.}
\label{fig2}
\end{center}
\end{figure}

\begin{figure}
\begin{center}
\includegraphics[width=7.8cm]{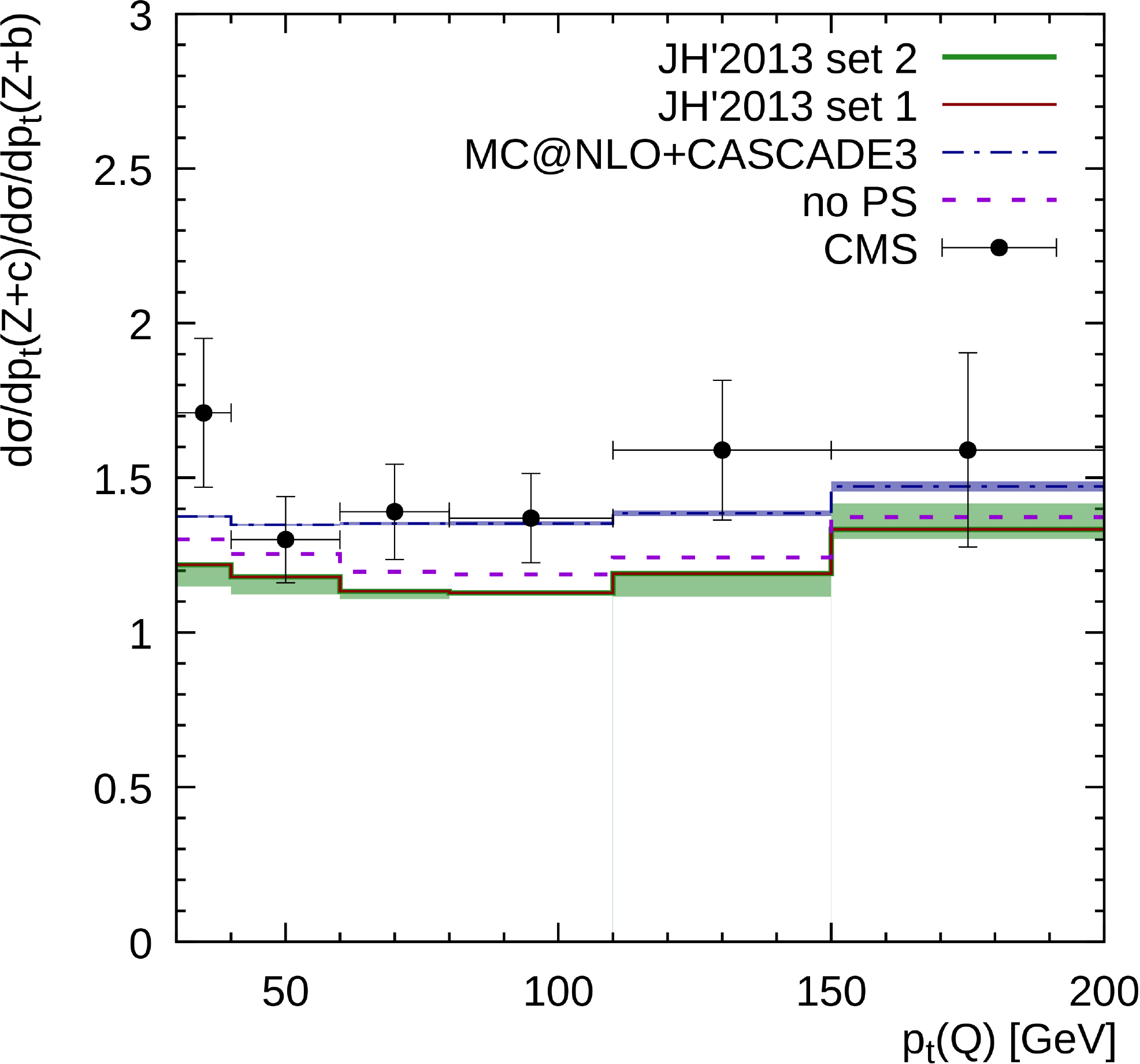}
\includegraphics[width=7.8cm]{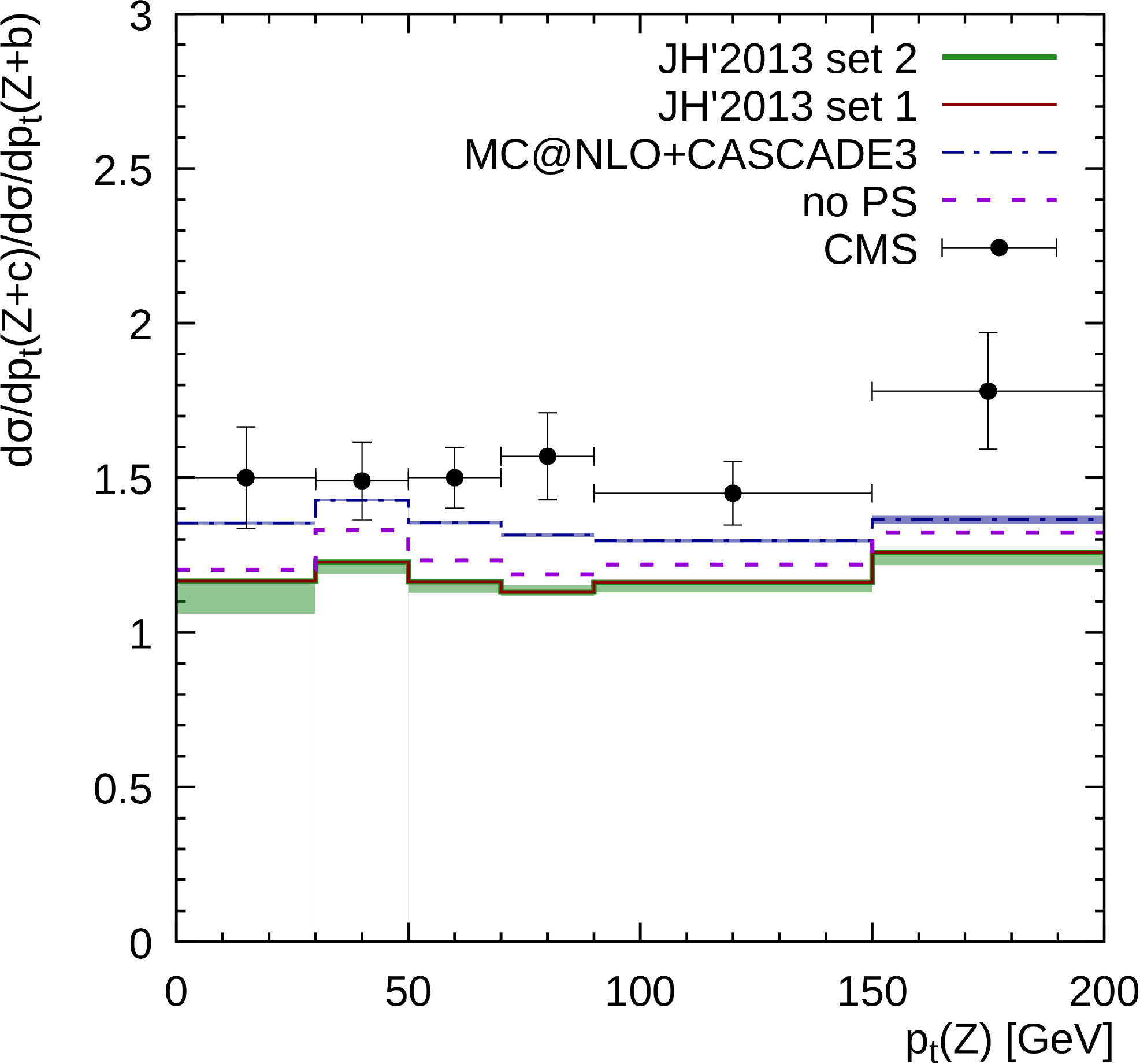}
\caption{The relative production rate $\sigma(Z + c)/\sigma(Z + b)$ as functions of heavy flavor jet (left panel) 
and $Z$ boson (right panel) transverse momenta at $\sqrt{s}= 13$~TeV. 
The notations are the same as in Fig.~1. 
The experimental data are from CMS\cite{7}.}
\label{fig3}
\end{center}
\end{figure}

We start from differential cross sections of associated $Z + c$ production at the LHC.
Results of our calculations are presented in Figs.~\ref{fig1} --- \ref{fig3}
in comparison with the CMS data\cite{5,6} taken at $\sqrt{s} = 8$ and $13$~TeV. 
The kinematical cuts and jet reconstructing parameters were taken the same as in corresponding
experimental analyses (we summarized them in Table~1). 
The shaded bands represent the theoretical uncertainties of our calculations.
To estimate the latter in the \textsc{pegasus} 
simulation we have used auxiliary "$+$" and "$-$" TMD gluon densities in a proton
instead of default ones when
calculating the off-shell gluon-gluon fusion subprocess~(\ref{ggZ}).
These two sets refer to the varied hard scales in the strong coupling
$\alpha_s$ 
in the off-shell amplitude: "$+$" 
stands for $2\mu_R$, while "$-$" refers to $\mu_R/2$.
This was done to preserve the intrinsic consistency of CCFM-based calculations 
(see\cite{Hautmann:2013} for more information).
For the quark-induced subprocesses~(\ref{qqtZ}) --- (\ref{qg5}) we just vary
the hard scales around its 
default value between halved and doubled magnitude, as it usually done.
We find that the measured $Z + c$-jet production 
cross sections 
are reasonably well reproduced by the \textsc{pegasus} calculations
(within the theoretical and experimental uncertainties),
although some underestimation of the CMS data taken at $\sqrt s = 8$~TeV
is observed at low $p_T(c)$ and large $p_T(Z)$.
A similar description of the $8$~TeV
data is achieved in the traditional (collinear)
NLO pQCD evaluations, as one can see in Fig.~1.
At the same time it is worth pointing out that the two analyses~\cite{5,6} used different techniques for the experimental charm jet identification. Namely, in the $8$~TeV measurement\cite{5} several 
methods were utilized for charm identification, including those based on the presence of a muon in the jet 
or the reconstruction of $D^\pm$ or $D^*(2010)^\pm$ meson exclusive decays. The $13$~TeV 
measurement benefited however from dedicated machine learning methods developed for identification 
of charm jets\cite{6}. 
Interestingly, unlike the \textsc{pegasus} predictions, the \textsc{cascade} results 
tend to overestimate the $8$~TeV data. The predictions of both TMD-based 
approaches
as well as NLO pQCD ones
are close to each other at $\sqrt s = 13$~TeV.
The calculated contribution from the DPS production mechanism is 
small for both considered energies and can 
play a role at low transverse momenta only.

We find that the \textsc{pegasus} predictions substantially depend on the TMD gluon used,
as one can see in Figs.~\ref{fig1} and~\ref{fig2}.
This can be explained by the fact that the off-shell gluon fusion 
subprocess~(\ref{ggZ}) plays an essential role in the considered kinematical region. 
Our calculations show that the CMS data taken at $\sqrt s = 8$~TeV are better 
described by JH'2013 set 1 gluon density (except for the first bin at $25<p_T<30$~GeV).
Moreover, these predictions practically
coincide with the corresponding results of the NLO pQCD calculations.
The quark-induced contributions~(\ref{qqtZ}) --- (\ref{qg5}) become important 
at high transverse momenta, where the typical $x$ values are large, 
that supports using DGLAP dynamics for these subprocesses.
Of course, these subprocesses should be taken into account to describe the 
data in the whole $p_T$ range.

To investigate the effects originating from initial and/or final state 
parton showers in the scheme implemented into the \textsc{pegasus} tool, 
we show separately
the results obtained at the parton level, that
corresponds to the previous calculations\cite{18,19}.
We find that simulation of parton showers 
leads to some decrease of the calculated cross sections.
However, the estimated effect is almost negligible and lies mostly 
within the bands of theoretical uncertainties.

Concerning the relative $\sigma(Z + c)/\sigma(Z + b)$ production rate, 
we find that the \textsc{pegasus} tends to underestimate recent CMS data,
whereas the \textsc{cascade} tool gives better description of the latter. In fact, there is only 
some underestimation of this ratio at the large transverse momenta of $Z$ boson, see Fig.~3. 
The observed difference between the \textsc{pegasus} and \textsc{cascade} predictions can be explained by different 
treatment in the two approaches: in \textsc{pegasus} one always has a heavy flavor jet in 
the final state, while \textsc{cascade} operates with a sample containing $Z$+any jets, 
from which only events having heavy flavor jets after showering are considered~\cite{JungZb2021}. 
However, the two methods are both compatible with the data within $\sim2\sigma$.

\begin{figure}
\begin{center}
\includegraphics[width=7.8cm]{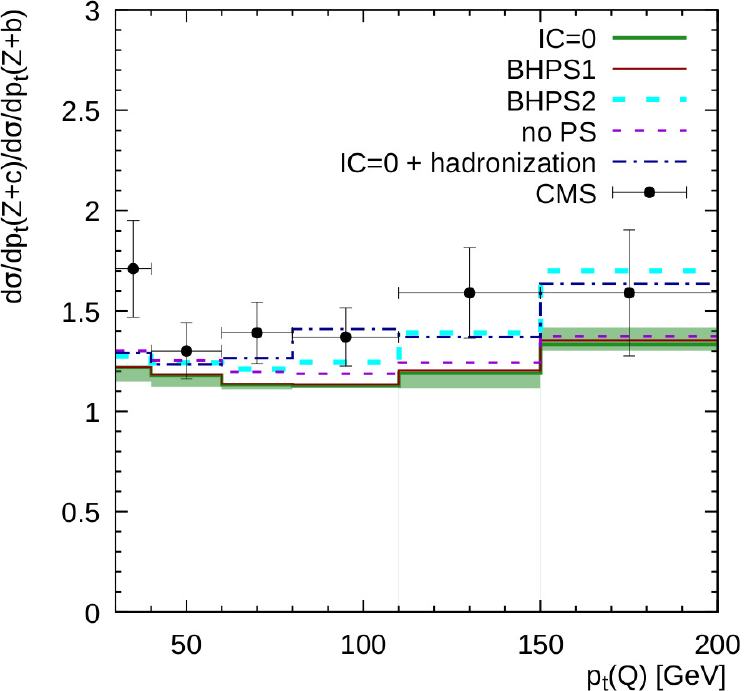}
\includegraphics[width=7.8cm]{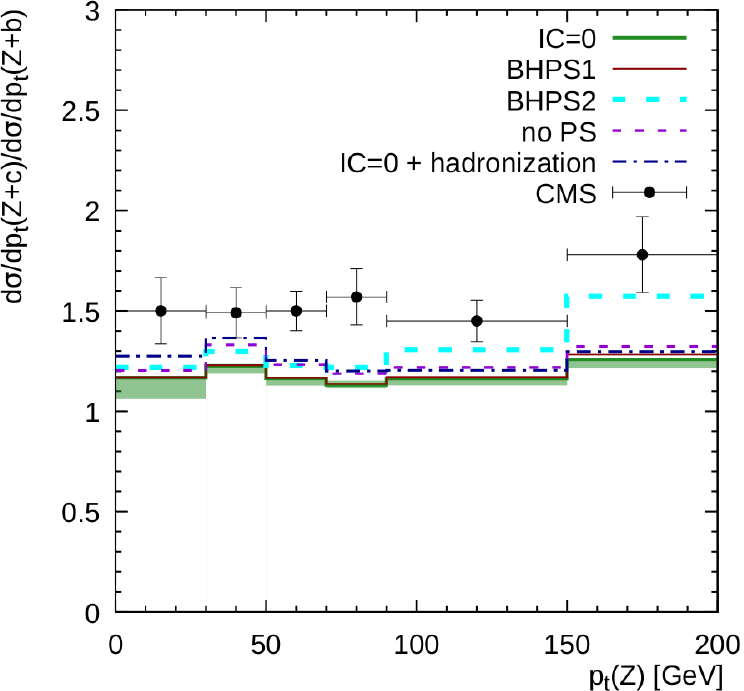}
\caption{The relative production rate $\sigma(Z + c)/\sigma(Z + b)$ as functions of heavy flavor jet (left panel) 
and $Z$ boson (right panel) transverse momenta 
calculated at $\sqrt{s}= 13$~TeV for different IC scenarios
with \textsc{pegasus}. Also results calculated without PS and with hadronization effects are shown. The experimental data are from CMS\cite{7}.}
\label{fig6}
\end{center}
\end{figure}

Now we turn to the next point of our study
connected with the investigation of heavy quark densities in a proton.
In fact, the production of vector bosons 
accompanied by heavy flavor jets in $pp$ collisions at the LHC can be considered as an
additional tool to study the quark and gluon densities in a proton.
As it is shown in \cite{18,10}, the sensitivity of $p_t$ spectra of 
prompt photons, $Z$-bosons and $c$-jets produced in 
$pp\rightarrow \gamma/Z + c + X$ processes at LHC energies to
different proton PDFs without the inclusion of the IC component
is very small, it is about a few percents. 
It would be very interesting to study the similar sensitivity to PDFs, which
include the IC contribution.   
To investigate these IC effects in more detail, we repeat
the calculations of associated $Z + c$-jet production cross sections
using the CT14 (NNLO) parton densities 
adopted for BHPS1 (corresponding to the IC 
probability $w_{\rm IC}^{\rm max}=1$\%) 
and BHPS2 (with $w_{\rm IC}^{\rm max}=3.5$\%) scenarios.
Results of our calculations, performed with Monte-Carlo generator \textsc{pegasus}, are shown in Figs.~1 and 2.
We find that the IC component is almost undetectable 
in the kinematical conditions of the CMS experiments even at high transverse 
momenta. Moreover, even being estimated within the BHPS2 scenario, IC signal 
lies within the bands of scale uncertainties of our calculations.
This agrees with the earlier results\cite{10,18,28,29,Bednyakov:2014,Beauchemin:2014},
where it was shown that the IC signal can be sizable in the forward rapidity region, $|y| \geq 1.5$.
The IC effect could be more visible, especially at large 
transverse momenta (about of $100$~GeV and higher)
in the relative production rate $\sigma(Z + c)/\sigma(Z + b)$
since most of theoretical uncertainties cancels out in this ratio (see Fig.~\ref{fig6}).

However, it is known that hadronization effects can result in a significant 
decreasing of the $Z$+heavy flavor jet production cross sections at least in some 
kinematical regions~\cite{JungZb2021}. We check the effect by applying the hadronization
effects in our \textsc{pegasus} calculations for the $\sigma(Z + c)/\sigma(Z + b)$ cross section 
ratios measured by CMS~\cite{7}. Our results (Fig.~4) show that the hadronization 
corrections for $Z+c$ production are essentially smaller than for $Z+b$ production. 
This results in the increasing of the cross section ratio, especially at large 
transverse momenta of the heavy quark jet (thus leading to better agreement with the data). 
This agrees with results of~\cite{JungZb2021}, obtained with \textsc{cascade}. 
So the IC effects can be in fact hidden by the hadronization effects, at least for $p_T(Q)$ differential cross sections.

The following simple argument can be also useful for further IC studies.
Assuming that the IC distribution peaks at $x\sim 0.5$,
one can expect that the distribution in Feynman 
variable $x_F = 2 p_z/\sqrt{s}$\cite{10,Brodsky_Higgs}
would generally follow the initial quark density
thus giving rise to an enhancement of the cross sections at large $x_F$ values
even in a specific kinematical region.
Based on this point, we have calculated the cross section of 
$Z + c$ production as a function of $x_F$
using the \textsc{pegasus} tool.
The results of our evaluations are shown in Fig.~5.
The kinematical cuts applied are given in the last column of Table~1. 
Note that here we limit ourselves to $x_F<0.6$ to control statistical uncertainties. 
We find that, even for the quite conservative 
IC fraction, the predictions of BHPS1 scenario starts to lie over the uncertainty 
band of a null hypothesis at $x_F\gtrsim 0.15$. 
It can illustrate qualitatively the kinematical region, where the IC signal
can be visible. 


\begin{figure}
\begin{center}
\includegraphics[width=8cm]{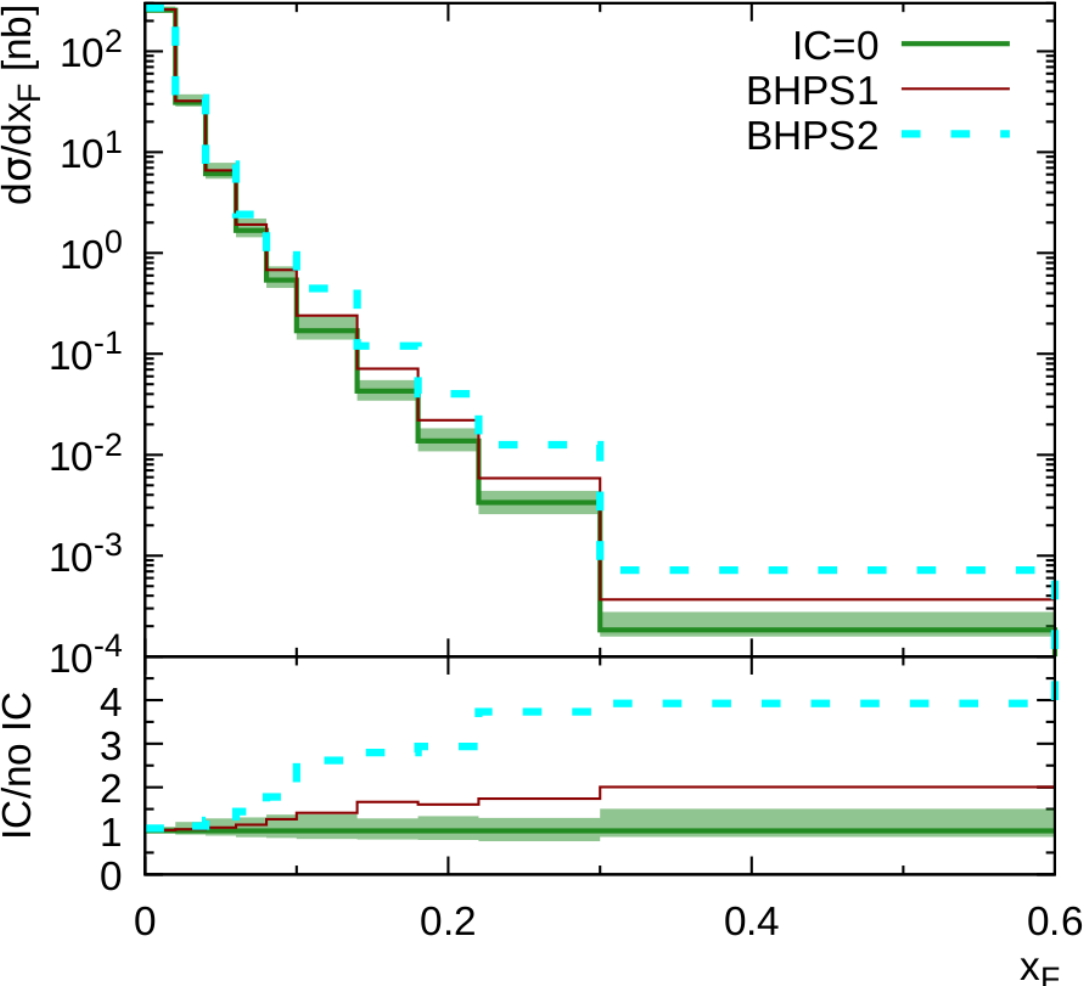}
\includegraphics[width=7.8cm]{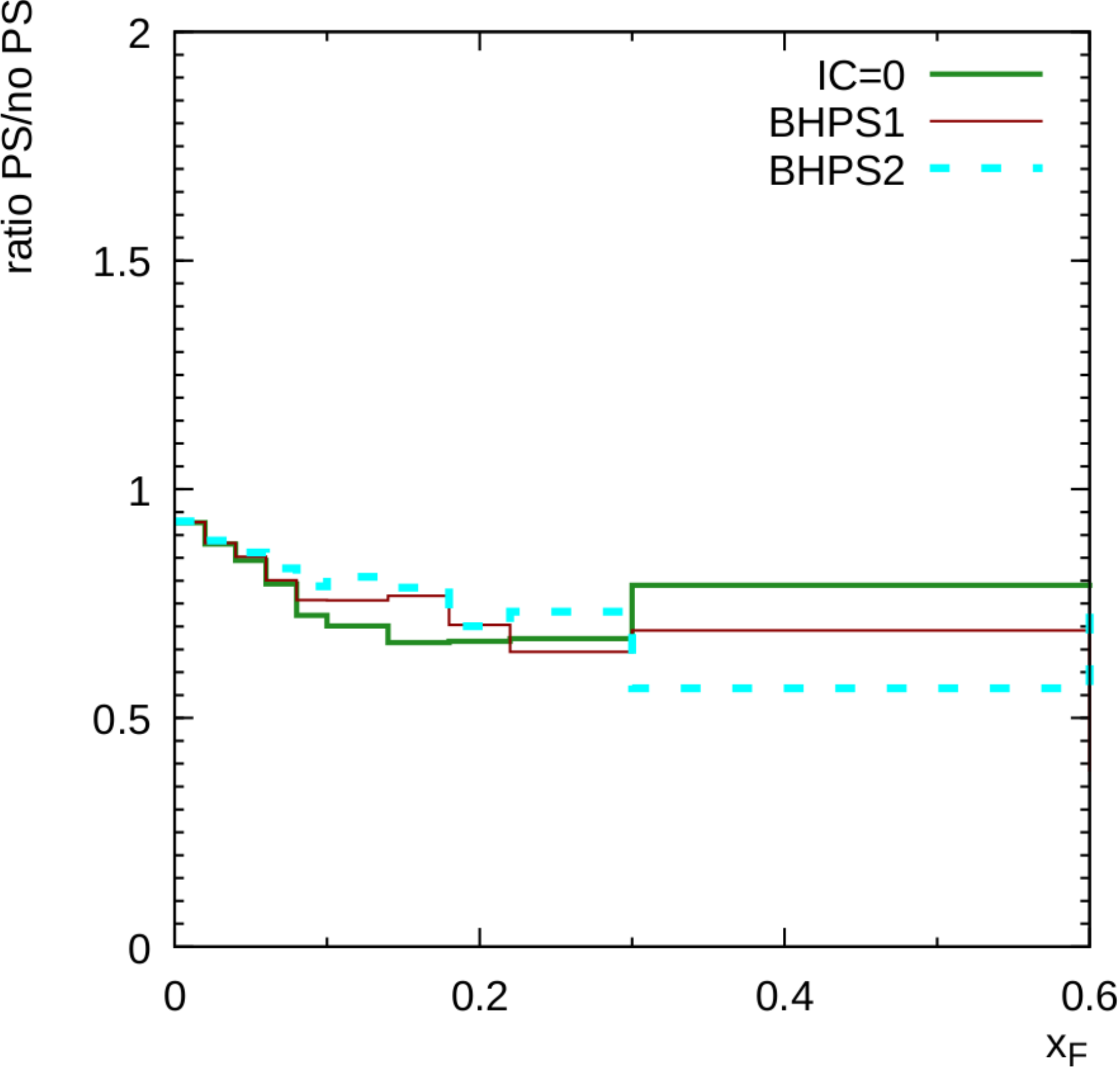}
\caption{The differential cross sections of associated $Z + c$ production
in $pp$ collisions at $\sqrt s = 13$~TeV for different intrinsic charm parametrization and their ratios to the zero IC scenario
calculated using \textsc{pegasus} tool as a function of Feynman variable $x_F$ (left).
Right: The ratio of the cross sections calculated with and without PS for the different parametrizations.}
\label{fig8}
\end{center}
\end{figure}

\section{Conclusion}\indent

We have considered the production of $Z$ bosons associated with heavy (charm and beauty) 
jets at the LHC energies using two TMD-based scenarios. 
The first approach employs the CCFM gluon evolution and has been
implemented in the Monte-Carlo event generator \textsc{pegasus}.
In this scenario, the heavy quarks are always produced in the hard partonic 
scattering subprocesses (\ref{ggZ}) --- (\ref{qg5}).
The second scheme is based on the PB approach implemented into the Monte-Carlo event 
generator \textsc{cascade}.
The traditional NLO pQCD calculations 
were done also using the standard \textsc{MadGraph5}\_a\textsc{mc@nlo} tool.


The main goal of this paper is to check the sensitivity of our results to
inputs used by calculations, namely: two different TMD gluon distributions (JH'2013 set 1 and JH'2013 set 2), 
the contribution of the parton showers,
the contribution of double parton scattering, different schemes of the QCD 
calculation, the QCD scale uncertainty, different sets of the conventional PDFs including also
the intrinsic charm contribution. 
We find that
there is a sensitivity of transverse momentum distributions of the $Z$-boson and $c$-jet to different TMD
gluon densities in a proton. In particular, the JH'2013 set 1 gluon leads to a small increase of the $p_T$ spectra 
of both $Z$ boson and $c$-jet about a few percents in the whole rapidity region.
The $p_T$ spectra of $c$-jet or $Z$ boson after inclusion of PS, DPS and
IC are changed also by about a few percents at $|y|\leq$ 2.4.
The sensitivity of all our results to the QCD scale uncertainty is about 10 percent in the whole rapidity range.
We show also that the contribution from the double parton scattering
mechanism is rather small and can play a role at low transverse momenta only.
It has been shown that the IC contribution to the ratio $\sigma(Z + c)/\sigma(Z + b)$ as a function of 
heavy flavor jet transverse momentum integrated over the rapidity $|y|\leq$ 2.4 can be hidden by the hadronization effects.
We have illustrated qualitatively that the IC signal can be visible in the $x_F$ distribution of $c$-jet at $x_F > $ 0.1,
which roughly correponds to large values of $p_T(Q)$ and the rapidity range $|y|>$ 1.5.      
It has been found that both considered TMD approaches
provide a more or less consistent description of recent
experimental data on the $Z$ boson and $c$-jet transverse momentum 
distributions.
This can be seen from a direct comparison between the \textsc{pegasus} and \textsc{cascade}
predictions and CMS data collected at $\sqrt s = 8$ and $13$~TeV.
Similar agreement with the data is achieved also with 
\textsc{MadGraph5}\_a\textsc{mc@nlo}.
However, the Monte-Carlo generator \textsc{cascade} provides a better 
description of the relative $\sigma(Z + c)/\sigma(Z + b)$ production rate,
that is connected with the different jet 
production mechanisms implemented into the \textsc{cascade} and \textsc{pegasus}.
 


\section*{Acknowledgements}

We thank S.P.~Baranov and A.~Bermudez Martinez for important comments and remarks.
We are obliged a lot to H.~Jung and S.~Taheri Monfared for providing the \textsc{CASCADE} and \textsc{MadGraph5}\_a\textsc{mc@nlo} results and allowing us to use them in this work.
A.V.L. and M.A.M. are grateful to
DESY Directorate for the support in the framework of
Cooperation Agreement between MSU and DESY on 
phenomenology of the LHC processes
and TMD parton densities.
M.A.M. was also supported by the grant of the Foundation for the Advancement of 
Theoretical Physics and Mathematics “BASIS” 20-1-3-11-1.

\end{document}